\documentclass[preprint,12pt]{elsarticle}

\usepackage{amssymb}
\usepackage{amsmath}
\usepackage{graphicx}
\usepackage{booktabs}
\usepackage{longtable}
\usepackage{algorithm}
\usepackage{algpseudocode}
\usepackage{url}
\usepackage{hyperref}

\hypersetup{
    colorlinks=true,
    linkcolor=blue,
    citecolor=blue,
    urlcolor=blue,
    filecolor=blue
}

\journal{Array}

\begin{document}

\begin{frontmatter}

\title{Multimodal Fusion at Three Tiers: Physics-Driven Data Generation and Vision-Language Guidance for Brain Tumor Segmentation}

\author{Mingda Zhang}
\ead{mingdazhang@acm.org}
\address{Software School, Yunnan University, Kunming 650504, Yunnan, China}

\begin{abstract}
Accurate brain tumor segmentation is crucial for neuro-oncology diagnosis and treatment planning. Deep learning methods have made significant progress, but automatic segmentation still faces challenges. These include tumor morphological heterogeneity and complex three-dimensional spatial relationships. This paper proposes a three-tier fusion architecture that achieves precise brain tumor segmentation. The method processes information progressively at the pixel, feature, and semantic levels. At the pixel level, physical modeling extends magnetic resonance imaging (MRI) to multimodal data. This includes simulated ultrasound and synthetic computed tomography (CT). At the feature level, the method achieves Transformer-based cross-modal feature fusion through multi-teacher collaborative distillation. The mechanism integrates three expert teachers ($T_{\text{MRI}}$, $T_{\text{US}}$, $T_{\text{CT}}$). At the semantic level, clinical textual knowledge generated by GPT-4V transforms into spatial guidance signals. This transformation uses Contrastive Language-Image Pre-training (CLIP) contrastive learning and Feature-wise Linear Modulation (FiLM). These three tiers work together to form a complete processing chain. The chain spans from data augmentation to feature extraction to semantic guidance. We validated the method on the Brain Tumor Segmentation (BraTS) 2020, 2021, and 2023 datasets. The model achieves average Dice coefficients of 0.8665, 0.9014, and 0.8912 on the three datasets, respectively. The 95\% Hausdorff Distance (HD95) reduces by an average of 6.57 millimeters compared to the baseline. This method provides a new paradigm for precise tumor segmentation and boundary localization.
\end{abstract}


\end{frontmatter}

\section{Introduction}

Magnetic resonance imaging (MRI) segmentation of brain tumors is fundamental for neuro-oncology diagnosis and treatment planning. Gliomas present significant challenges due to their infiltrative growth patterns, morphological heterogeneity, and multiple sub-regions~\cite{menze2015multimodal,bakas2017advancing}. Different sub-regions have different prognostic significance. Contemporary imaging protocols typically acquire multiple MRI sequences. These include T1-weighted, contrast-enhanced T1-weighted (T1ce), T2-weighted, and fluid-attenuated inversion recovery (FLAIR)~\cite{menze2015multimodal}. Each sequence provides complementary tissue characterization information. Recent deep learning methods have made remarkable progress, particularly three-dimensional convolutional neural networks (CNNs)~\cite{cicek20163d,milletari2016vnet} and U-Net~\cite{ronneberger2015unet} variants with encoder-decoder architectures. Fully automatic networks~\cite{isensee2021nnu} have shown application potential in this field. However, these methods still face limitations. They struggle with blurred tumor boundaries, morphological heterogeneity, and multimodal information integration.

Attention mechanisms provide a theoretical framework for feature selection and fusion. They work through learnable information bottlenecks that selectively enhance task-relevant information while suppressing redundant information. Channel attention~\cite{hu2018squeeze} identifies important feature dimensions. Spatial attention~\cite{woo2018cbam} localizes key regions. Self-attention mechanisms model long-range dependencies~\cite{wang2021transbts,chen2021transunet}. These mechanisms show significant advantages when processing brain tumor images with target-background imbalances and infiltrative boundaries~\cite{oktay2018attention,roy2018recalibrating}. They effectively capture weak lesion signals and significantly improve segmentation accuracy. However, existing methods often apply attention mechanisms in isolation at a single stage. They fail to explore synergistic effects across different levels. This includes pixel-level multimodal generation, feature-level cross-modal alignment, and semantic-level text guidance. The result is relatively independent processing at each level.

Multi-scale feature fusion is grounded in scale-space theory. This theory indicates that different image structures have optimal representations at different scales. In brain tumor segmentation, coarse-scale features capture macroscopic lesion distribution. Fine-scale features preserve boundary details and internal heterogeneous structures. Encoder-decoder architectures extract multi-scale representations and use skip connections~\cite{zhou2018unet++,hatamizadeh2022unetr}. These connections fuse shallow details with deep semantics, achieving both global consistency and local precision. Skip connections provide short paths for gradient propagation. They effectively alleviate gradient vanishing and avoid information bottlenecks. However, existing methods mainly focus on single-level visual feature processing. They typically use simple channel concatenation or weighted fusion. They lack systematic mechanisms that integrate different abstraction levels. This single-level processing cannot fully exploit information complementarity at different levels. It limits the model's ability to understand complex lesion features.

To overcome the limitations of pure visual feature processing, introducing high-level semantic knowledge has become critical. Contrastive Language-Image Pre-training (CLIP) models~\cite{radford2021learning} learn cross-modal semantic representations through contrastive learning. They maximize similarity of matching pairs while minimizing similarity of mismatched pairs. This paradigm demonstrates advantages in medical image segmentation. It establishes semantic alignment between visual patterns and radiological terminology~\cite{huang2021gloria}. It encourages discriminative feature boundaries between different pathological tissues. It also provides visual-semantic prior knowledge from large-scale pre-training~\cite{dosovitskiy2020image}. However, applying CLIP to three-dimensional medical image segmentation faces challenges. These include dimensional mismatch between two-dimensional (2D) pre-trained models and three-dimensional (3D) volumetric data. There is also domain shift between natural and medical images~\cite{zhang2023biomedclip}. Additionally, task conversion from image-level understanding to pixel-level segmentation poses difficulties. Existing work~\cite{zhao2024large,koleilat2024medclipsam} lacks systematic mechanisms. They cannot transform CLIP's semantic understanding into spatial guidance for 3D segmentation. They fail to establish complete mapping paths from textual concepts to spatial attention allocation.

To address these challenges, this paper proposes a three-tier fusion architecture. It organically integrates pixel-level multimodal generation, feature-level cross-modal fusion, and semantic-level concept guidance. The main contributions include:

1. We propose a three-tier fusion architecture that integrates physics-driven multimodal generation~\cite{li2024diffusion}, Transformer-guided cross-modal alignment, and CLIP-based semantic guidance. The integration occurs at the pixel, feature, and semantic levels.

2. We design an asynchronous progressive multi-teacher distillation mechanism~\cite{ahmad2024multi,hossain2024enhancing}. It adaptively integrates knowledge from three expert teachers ($T_{\text{MRI}}$, $T_{\text{US}}$, $T_{\text{CT}}$) and one semantic teacher ($T_{\text{GPT-4V}}$). The integration uses dynamic weight adjustment.

3. We construct a concept-driven semantic guidance mechanism. It uses CLIP contrastive learning and GPT-4V descriptions~\cite{brin2024comparing}. The mechanism transforms medical concepts into spatial guidance through Feature-wise Linear Modulation (FiLM) and attention generation~\cite{zhang2024multi}.

\section{Related Work}

Brain tumor segmentation technology development has important implications for neuro-oncology diagnosis and treatment. Gliomas present challenges due to their infiltrative growth, morphological heterogeneity, and multiple sub-regions. Different sub-regions have different clinical significance. The BraTS challenge~\cite{bakas2018identifying,baid2021rsna} established standardized evaluation protocols. These protocols cover whole tumor (WT), tumor core (TC), and enhancing tumor (ET). The challenge has promoted rapid development of deep learning methods. However, precise segmentation still faces challenges. These include blurred tumor boundaries and internal heterogeneity. Effective multimodal MRI integration remains difficult. There is also a lack of mechanisms to explicitly introduce clinical knowledge into the segmentation process.

\subsection{Multi-Scale Feature Fusion}

The encoder-decoder architecture based on U-Net~\cite{ronneberger2015unet} established the foundational paradigm for medical image segmentation. It achieved this through symmetric structure and skip connections. Subsequent research has evolved along three main directions. Dimensional extension methods such as 3D U-Net~\cite{cicek20163d} and V-Net~\cite{milletari2016vnet} process volumetric data to capture complete spatial relationships. Adaptive optimization methods like nnU-Net~\cite{isensee2021nnu} achieve optimal performance through automatic configuration. Multi-scale enhancement methods strengthen cross-scale feature fusion. These include UNet++~\cite{zhou2018unet++}, Multi-Scale Reverse Attention Module (MSRAM)~\cite{zeng2023msram}, and Region-Attention Fusion Network (RFTNet)~\cite{jiao2024rftnet}. Multi-scale concepts have also been applied to other tasks. Multi-Scale Feature Network (MsfNet)~\cite{song2021msfnet} and Multi-Scale Feature You Only Look Once (MSFYOLO)~\cite{song2022msfyolo} achieve good results in small object detection. Cross-scale Wavelet Transform Network (CWT-Net)~\cite{jia2024cwt} combines cross-scale wavelet transform with Transformers. Recent generative models have also shown potential. Diffusion methods~\cite{wolleb2022diffusion,nie2025diffbts,wu2024fcfdiff} generate high-quality segmentation through conditional denoising. Multimodal generation techniques~\cite{kebaili2024multi,zarif2024enhanced} improve generalization. However, these methods mainly focus on single-level feature processing. They use simple concatenation or weighted fusion. They lack systematic mechanisms that integrate pixel-level data augmentation, feature-level cross-modal alignment, and semantic-level concept guidance. This makes it difficult to fully exploit information complementarity at different abstraction levels.

The three-tier fusion architecture proposed in this paper systematically deploys multi-scale processing mechanisms. It operates at the pixel, feature, and semantic levels. The architecture establishes a complete information flow from low-level data to high-level concepts through cross-level collaborative optimization.

\subsection{Attention Mechanisms}

Attention mechanisms have evolved from single-dimensional to multi-dimensional, and from local to global. Early methods achieve information bottleneck optimization through data-driven weight allocation. These include Squeeze-and-Excitation Network (SE-Net)~\cite{hu2018squeeze}, Convolutional Block Attention Module (CBAM)~\cite{woo2018cbam}, and spatial and channel Squeeze and Excitation (scSE)~\cite{roy2018recalibrating}. Medical-specific attention methods such as Attention U-Net~\cite{oktay2018attention} improve segmentation accuracy by suppressing irrelevant regions. The Transformer architecture brought paradigm shifts. Transformer-based Brain Tumor Segmentation (TransBTS)~\cite{wang2021transbts} and TransUNet~\cite{chen2021transunet} introduced self-attention mechanisms. They broke through local receptive field limitations. Hybrid architectures combine CNN's local feature extraction with Transformer's global dependency modeling. These include U-Net Transformer (UNETR)~\cite{hatamizadeh2022unetr} and Swin U-Net Transformer (SwinUNETR)~\cite{hatamizadeh2022swin}. Methods incorporating domain priors attempt to use clinical knowledge to guide attention allocation. Examples include Swin Transformer-based Brain Tumor Segmentation (SwinBTS)~\cite{jiang2022swinbts} and Clinical Knowledge-Driven TransBTS (CKD-TransBTS)~\cite{lin2023ckd}. However, these methods often apply attention in isolation at single stages. They lack systematic deployment across pixel-level multimodal generation, feature-level cross-modal alignment, and semantic-level text guidance. This results in relatively independent processing at each level.

This paper systematically deploys attention mechanisms at multiple levels. These include cross-modal alignment in feature fusion, feature refinement in the encoder, and spatial guidance at the semantic level. This forms a complete attention flow from low-level features to high-level semantics.

\subsection{Vision-Language Models}

Vision-language pre-trained models have opened new directions for medical image analysis. CLIP~\cite{radford2021learning} and Vision Transformer (ViT)~\cite{dosovitskiy2020image} establish strong zero-shot transfer capabilities. However, they suffer from domain shift between natural and medical images. Global-Local Representations for Images using Attention (GLoRIA)~\cite{huang2021gloria} and BiomedCLIP~\cite{zhang2023biomedclip} narrow the domain gap through medical domain adaptation. They achieve excellent performance in image-level tasks. Recent work explores weakly supervised segmentation through text prompts. Examples include Zhao et al.~\cite{zhao2024large} and MedCLIP-SAM~\cite{koleilat2024medclipsam}. However, these are mainly limited to 2D images. Progress in multimodal generative artificial intelligence (AI) shows that text descriptions can guide MRI synthesis~\cite{wang2024toward}. GPT-4V demonstrates capabilities in radiological analysis~\cite{brin2024comparing}. However, current research faces key challenges. There are dimensional gaps between 2D pre-trained models and 3D medical volumes. Domain shift exists between natural and medical images~\cite{zhang2023biomedclip}. Systematically transforming contrastive learning advantages into spatial guidance mechanisms remains difficult. Existing work~\cite{zhao2024large,koleilat2024medclipsam} lacks complete design. It cannot transform CLIP's semantic understanding into spatial guidance for 3D segmentation.

This paper resolves dimensional mismatch through a 3D-2D semantic bridging mechanism. It extracts representative slices from three orthogonal planes and fuses multi-view semantic information. We design a cross-modal semantic guidance mechanism. It transforms CLIP's image-level semantic understanding into pixel-level spatial guidance signals. This transformation uses FiLM and spatial attention generation~\cite{zhang2024multi}.

\section{Method}

\begin{figure}[htbp]
  \centering
  \includegraphics[width=\textwidth]{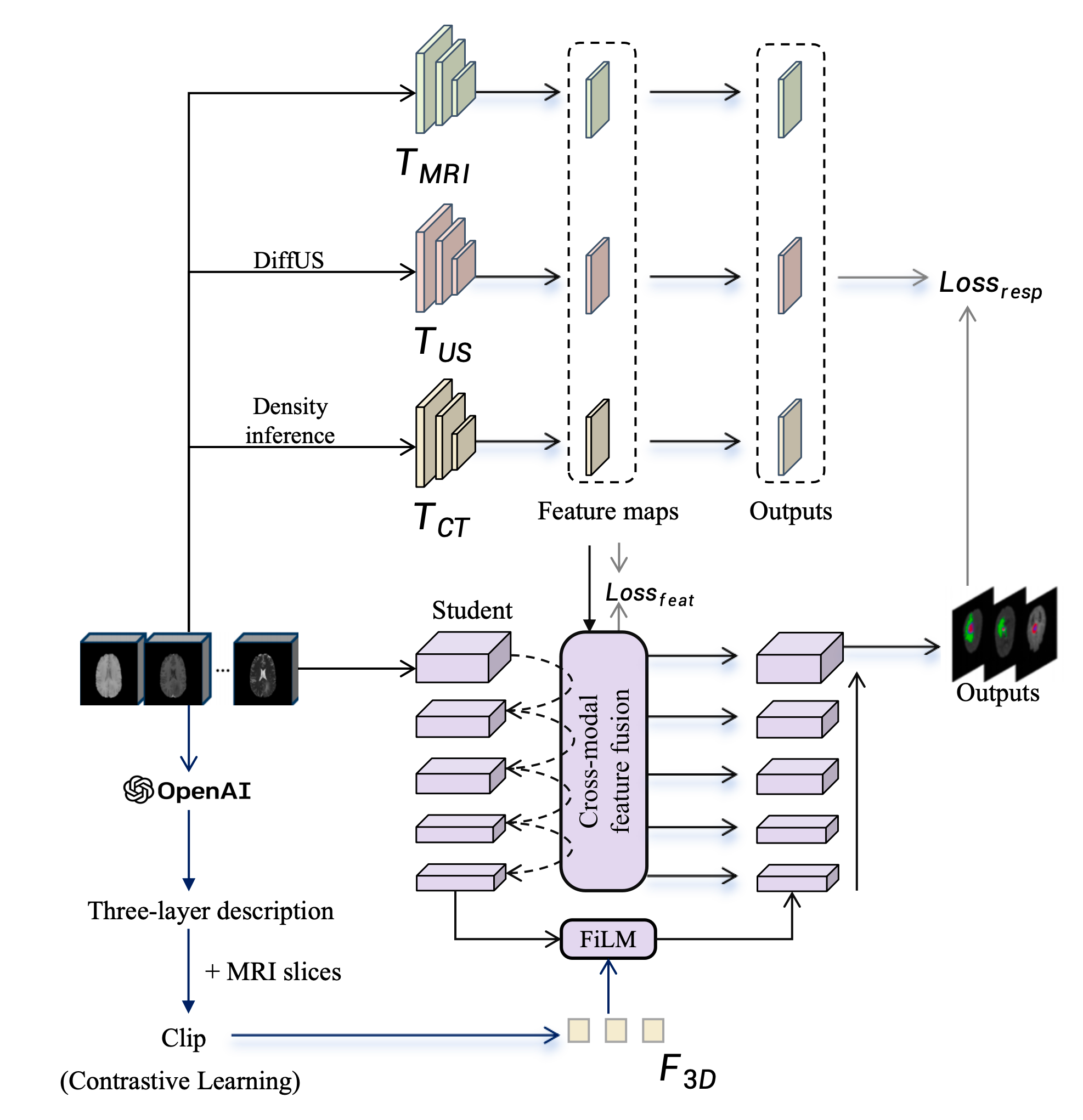}
  \caption{Overall framework of the three-tier fusion architecture. The architecture processes MRI input through three parallel pathways: $T_{\text{MRI}}$ (original MRI teacher), $T_{\text{US}}$ (simulated ultrasound teacher via DiffUS), and $T_{\text{CT}}$ (synthetic CT teacher via density inference). Feature maps and outputs from the three teachers guide the student network through cross-modal feature fusion. GPT-4V generates clinical descriptions from MRI slices. CLIP contrastive learning transforms these into semantic guidance $F_{\text{3D}}$. The FiLM (Feature-wise Linear Modulation) module modulates student features with semantic information. This enables precise segmentation through multi-teacher distillation ($\mathcal{L}_{\text{resp}}$ and $\mathcal{L}_{\text{feat}}$).}
  \label{fig:framework}
\end{figure}

This paper proposes a multimodal brain tumor segmentation method based on a three-tier fusion architecture that systematically integrates pixel-level, feature-level, and semantic-level processing. As shown in Figure~\ref{fig:framework}, the method employs three parallel pathways: the MRI teacher $T_{\text{MRI}}$ processes original sequences, the ultrasound teacher $T_{\text{US}}$ uses simulated ultrasound from the DiffUS module, and the CT teacher $T_{\text{CT}}$ processes synthetic CT from density inference. Cross-modal feature fusion integrates knowledge from these expert teachers into the student network. Meanwhile, GPT-4V generates clinical descriptions from representative MRI slices, which CLIP transforms into semantic guidance $F_{\text{3D}}$ through contrastive learning. The FiLM module then injects this semantic information into the student network's features, achieving concept-to-space transformation. This processing chain—spanning data augmentation, feature extraction, and semantic guidance—simulates the cognitive process of radiologists combining visual observation with clinical knowledge.

\subsection{Pixel-Level Fusion: Multimodal Data Generation and Preprocessing}

Pixel-level fusion optimizes the original MRI data and generates complementary modalities. This provides a rich multi-source data foundation for subsequent feature extraction. After the four-channel MRI data enters the system, the data flows to three parallel paths. The physical rendering path generates simulated ultrasound images. The density inference path generates synthetic CT images. The preprocessing path optimizes the original MRI sequences.

The physical rendering path uses the DiffUS module to convert MRI to simulated ultrasound images. To establish the mapping between MRI signal intensity and ultrasound acoustic properties, this module first converts the MRI volume to an acoustic impedance representation:
\begin{equation}
Z_{\text{acoustic}} = f_{\text{MRI}\rightarrow Z}(I_{\text{MRI}}; \theta_{\text{mapper}})
\end{equation}
Here, $I_{\text{MRI}}$ represents the input MRI volume data. The function $f_{\text{MRI}\rightarrow Z}$ is a learnable mapping function implemented by a small convolutional neural network. The parameter $\theta_{\text{mapper}}$ represents the trainable parameters. The output $Z_{\text{acoustic}}$ is the acoustic impedance volume. This mapping process learns the nonlinear relationship between MRI signal intensity and tissue density. It enables different tissue types to receive reasonable acoustic properties.

After obtaining the acoustic impedance, the system generates B-mode ultrasound images through ray tracing algorithms. These algorithms simulate ultrasound wave propagation:
\begin{equation}
I_{\text{US}} = \text{RayTrace}(Z_{\text{acoustic}}) \otimes \text{PSF}_{\text{US}}
\end{equation}
The operator $\text{RayTrace}(\cdot)$ represents the ray tracing process. It simulates wave reflection, refraction, and attenuation at tissue interfaces. The symbol $\otimes$ denotes convolution. The function $\text{PSF}_{\text{US}}$ is the point spread function. It models depth-dependent blurring and speckle noise. This physics-driven generation captures strong echo features at tumor boundaries. It also captures heterogeneous echo patterns internally.

The density inference path generates synthetic CT through Bayesian mixture modeling:
\begin{equation}
p(I_{\text{CT}}|I_{\text{MRI}}) = \sum_{k=1}^{K} \pi_k \mathcal{N}(I_{\text{CT}}; \mu_k, \Sigma_k)
\end{equation}
Here, $K$ represents the number of tissue types. The weight $\pi_k$ is the mixture weight satisfying $\sum_{k=1}^{K}\pi_k=1$. The distribution $\mathcal{N}(\cdot; \mu_k, \Sigma_k)$ represents a Gaussian distribution. Statistical Parametric Mapping 12 (SPM12) estimates its parameters. This model expresses that each voxel's CT value comes from a probabilistic mixture of multiple tissue types.

\subsection{Feature-Level Fusion: Cross-Modal Feature Extraction and Multi-Teacher Collaborative Distillation}

The feature-level fusion layer transforms pixel-level generated heterogeneous data into unified high-level semantic representations. This stage addresses the fundamental challenge of fusing features from different imaging modalities. These features reside in their own independent representation spaces.

The cross-modal alignment network identifies semantically related modal correspondences by computing similarity between features. Despite different imaging mechanisms, semantically corresponding anatomical structures should have correlated feature representations. Cosine similarity serves as the metric:
\begin{equation}
A_{i,j} = \frac{f_i \cdot f_j}{\|f_i\| \|f_j\|}
\end{equation}
Here, $f_i$ and $f_j$ represent feature vectors from different modalities or spatial locations. The symbol $\cdot$ denotes inner product. The notation $\|\cdot\|$ denotes L2 norm. This normalized metric is unaffected by feature magnitudes. It enables fair comparison across modalities.

Based on the similarity matrix, the system achieves feature alignment and fusion through attention mechanisms:
\begin{equation}
F_{\text{aligned}} = \text{Softmax}(A) \cdot [F_{\text{MRI}}; F_{\text{US}}; F_{\text{CT}}]
\end{equation}
The notation $[\cdot;\cdot;\cdot]$ denotes feature concatenation. The function $\text{Softmax}(A)$ normalizes attention weights. This mechanism adaptively selects the most relevant features from each modality. It improves recognition accuracy for complex lesion patterns.

To systematically integrate expert knowledge from different modalities, this paper proposes an asynchronous progressive multi-teacher distillation mechanism. Knowledge distillation transfers dark knowledge from teacher models to student models. The mechanism constructs three expert teachers and one semantic teacher. The MRI teacher $T_{\text{MRI}}$ excels at identifying boundaries between tumor parenchyma and normal brain tissue. The CT teacher $T_{\text{CT}}$ focuses on density anomaly detection for calcification and hemorrhage. The ultrasound teacher $T_{\text{US}}$ provides precise contour localization. The GPT-4V semantic teacher $T_{\text{GPT-4V}}$ outputs structured knowledge consistent with clinical diagnostic logic. 

Knowledge transfer occurs through a dual-level distillation loss function:
\begin{equation}
\mathcal{L}_{\text{distill}} = \lambda_{\text{resp}} \sum_{i} \alpha_i \text{KL}(p_s \| p_{T_i}) + \lambda_{\text{feat}} \sum_{i} \alpha_i \|F_s - F_{T_i}\|_2^2
\end{equation}
The first term is response-level distillation. Here, $p_s$ represents the student's class probability distribution. The notation $p_{T_i}$ is the $i$-th teacher's output. The function $\text{KL}(\cdot\|\cdot)$ is Kullback-Leibler divergence measuring distribution difference. The weight $\alpha_i$ is the dynamic teacher weight. The second term is feature-level distillation. Here, $F_s$ and $F_{T_i}$ are student and teacher feature maps. The notation $\|\cdot\|_2^2$ measures Euclidean distance. This dual-level strategy guides both prediction outputs and intermediate representations.

The cross-modal feature fusion module serves as the bridge between teacher knowledge and student learning. Figure~\ref{fig:framework} illustrates how feature maps from $T_{\text{MRI}}$, $T_{\text{US}}$, and $T_{\text{CT}}$ aggregate through the cross-modal fusion mechanism. The FiLM module injects semantic guidance $F_{\text{3D}}$ from CLIP into the fusion process:
\begin{equation}
F_{\text{modulated}} = \text{FiLM}(F_{\text{fused}}, F_{\text{3D}}) = \gamma(F_{\text{3D}}) \odot F_{\text{fused}} + \beta(F_{\text{3D}})
\end{equation}
The functions $\gamma(\cdot)$ and $\beta(\cdot)$ are learned scaling and shifting functions. They modulate features based on semantic guidance. The symbol $\odot$ denotes element-wise multiplication.

Each encoding stage includes knowledge fusion nodes. These nodes dynamically integrate teacher knowledge through gating:
\begin{equation}
F_{\text{fused}}^{(l)} = F_{\text{student}}^{(l)} \odot \sigma(G^{(l)}) + F_{\text{teacher}}^{(l)} \odot (1 - \sigma(G^{(l)}))
\end{equation}
The notation $(l)$ denotes the network layer. The weight $G^{(l)}$ is a learnable gating weight. The function $\sigma(\cdot)$ is sigmoid activation. This allows adaptive balancing between student features and teacher knowledge.

\subsection{Semantic-Level Fusion: Cross-Modal Semantic Guidance Based on CLIP Contrastive Learning}

The lower portion of Figure~\ref{fig:framework} shows the semantic guidance pathway. It begins with GPT-4V generating structured clinical descriptions. These come from representative MRI slices extracted from three orthogonal planes. The CLIP model receives these textual descriptions and the corresponding image slices for contrastive learning. This produces the unified semantic representation $F_{\text{3D}}$. The FiLM module then injects this semantic signal into the student network. It transforms high-level medical concepts into spatial guidance for precise segmentation.

The CLIP model~\cite{radford2021learning} learned powerful cross-modal semantic representation capabilities through contrastive learning on large-scale image-text pairs, enabling it to establish semantic associations between clinical descriptions and imaging features while enhancing discriminative boundaries between different pathological tissues. Contrastive learning in CLIP learns a joint embedding space in which matching image-text pairs have high similarity while mismatched pairs have low similarity. The model achieves this by maximizing the cosine similarity of positive pairs and minimizing that of negative pairs. Contrastive learning uses the InfoNCE loss function:
\begin{equation}
\mathcal{L}_{\text{CLIP}} = -\frac{1}{N}\sum_{i=1}^{N}\log\frac{\exp(\text{sim}(v_i, t_i)/\tau)}{\sum_{j=1}^{N}\exp(\text{sim}(v_i, t_j)/\tau)}
\end{equation}
Here, $v_i$ and $t_i$ represent matched image and text features. The function $\text{sim}(\cdot,\cdot)$ represents cosine similarity. The parameter $\tau$ is the temperature controlling the distribution concentration. This loss function learns cross-modal semantic representations. It maximizes matching pair similarity and minimizes mismatched pair similarity.

To resolve the dimensional mismatch between CLIP and GPT-4V's 2D training and 3D medical volumes, this paper proposes a multi-view semantic guidance mechanism. The system extracts representative slices from three orthogonal anatomical planes (axial, coronal, sagittal) of the MRI volume, sends them to GPT-4V to generate layered medical text descriptions, and transforms these into semantic vectors via the CLIP text encoder. To form a unified semantic understanding of the entire 3D volume, the system fuses semantic vectors from the three orthogonal views:
\begin{equation}
F_{\text{3D}} = \frac{1}{3} \sum_{d \in \{\text{axial, coronal, sagittal}\}} F_s^d
\end{equation}
Here, $F_s^d$ represents the feature vector from the CLIP vision encoder for slices in direction $d$. It comes from the last layer's token output aggregating global semantic information. This multi-view fusion strategy leverages the complementarity of different anatomical planes. It avoids information bias from a single perspective.

To transform abstract semantic vectors into concrete guidance for 3D segmentation, this paper designs a cross-modal semantic guidance mechanism. Dedicated mapping networks project visual features and semantic vectors to a shared semantic space. The system then dynamically fuses them through gating:
\begin{equation}
G_{\text{text}} = \sigma(F_{\text{text\_mapped}})
\end{equation}
\begin{equation}
F_{\text{combined}} = F_{\text{vision\_mapped}} \odot G_{\text{text}} + F_{\text{text\_mapped}} \odot (1 - G_{\text{text}})
\end{equation}
The system processes $F_{\text{text\_mapped}}$ through sigmoid to obtain gating signal $G_{\text{text}}$. This indicates the degree to which text information should be adopted for each feature channel. This gating mechanism dynamically determines through learned weights. It decides the degree to which each position relies on visual information or text semantics.

Multi-head self-attention mechanisms further refine the fused features:
\begin{equation}
F_{\text{fused}} = \text{LayerNorm}(F_{\text{combined}} + \text{MultiHead}(F_{\text{combined}}))
\end{equation}
The function $\text{MultiHead}(\cdot)$ represents multi-head self-attention. It divides input features into multiple heads along the channel dimension. This captures long-range dependencies and global tumor structures. It models dependencies between any two positions.

A semantic attention generation mechanism transforms specific concepts in medical descriptions into explicit spatial localization signals:
\begin{equation}
M_{\text{semantic}} = \sigma(\text{Conv}_{1\times1\times1}([F_{\text{guided}}, s_{\text{spatial}}]))
\end{equation}
\begin{equation}
F_{\text{final}} = F_{\text{guided}} \odot M_{\text{semantic}} + F_{\text{guided}}
\end{equation}
Here, $s_{\text{spatial}}$ is the semantic vector expanded to match feature map dimensions. The function $\text{Conv}_{1\times1\times1}(\cdot)$ learns to generate spatial weights based on joint visual-text information. The mask $M_{\text{semantic}}$ modulates activation strength. Positions matching text descriptions receive stronger responses.

For the clinically critical ET and TC regions, the system deploys dedicated semantic enhancement modules:
\begin{equation}
Y_{\text{ET\_enhanced}} = Y_{\text{base\_ET}} \odot \sigma(f_{\text{conv}}(F_{\text{decoder1}}, F_{\text{fused}}))
\end{equation}
Here, $Y_{\text{base\_ET}}$ represents preliminary ET prediction. The features $F_{\text{decoder1}}$ are shallow decoder features preserving spatial details. The function $f_{\text{conv}}(\cdot, \cdot)$ is an attention generation network. This enables accurate localization of small-volume, high-complexity regions. It combines clinical text with visual features.

\section{Experiments}

\subsection{Experimental Setup}

Datasets. This paper validates the effectiveness of the method on three benchmark datasets from the BraTS Challenge: BraTS 2020, 2021, and 2023. Multiple international medical imaging institutions jointly organize the BraTS challenge. It provides multi-center, multi-device glioma MRI data collected from 19 medical centers worldwide~\cite{bakas2018identifying}. Each case includes four co-registered MRI sequences. These are T1-weighted (T1), contrast-enhanced T1-weighted (T1ce), T2-weighted (T2), and FLAIR sequences. Professional neuroradiologists perform layer-by-layer manual delineation based on unified annotation protocols. Labels include three hierarchical sub-regions. ET corresponds to enhancement regions in T1ce sequences, representing actively proliferating tumor parenchyma. TC includes ET, necrotic regions, and non-enhancing tumor. WT encompasses TC and surrounding edema regions with high T2/FLAIR signals. BraTS 2020 includes 369 training samples and 125 validation samples~\cite{bakas2018identifying}. BraTS 2021 expands to 1251 training samples and 219 validation samples~\cite{baid2021rsna}. BraTS 2023 maintains 1251 training samples while introducing more data from new medical centers~\cite{bakas2023brats}. The original data has a spatial resolution of $240 \times 240 \times 155$ voxels. Voxel spacing is approximately $1 \times 1 \times 1$ mm³.

Evaluation Metrics. This study uses the Dice Similarity Coefficient (Dice)~\cite{dice1945measures} and 95\% Hausdorff Distance (HD95)~\cite{karimi2020reducing} as evaluation metrics. The Dice coefficient calculates volumetric overlap: $\text{Dice} = 2|A \cap B|/(|A|+|B|)$. Here, $A$ represents the ground truth foreground and $B$ represents the predicted foreground. Values range from 0 to 1. HD95 first extracts surface voxel points from predictions and ground truth. It calculates the minimum Euclidean distance from each surface point to the other surface in both directions. It then merges bidirectional distances and takes the 95th percentile. This improves robustness by excluding the largest 5\% of outliers.

\subsection{Comparison with Existing Methods}

Table~\ref{tab:brats_all} systematically compares the proposed method with representative methods on BraTS 2020, 2021, and 2023. This provides a comprehensive evaluation of the three-tier fusion architecture. Comparison methods cover CNN architectures like nnU-Net~\cite{isensee2018nnu}, 3D U-Net~\cite{cicek20163d}, V-Net~\cite{milletari2016vnet}, and SegResNet~\cite{myronenko20183d}. They include Transformer methods such as TransBTS~\cite{wang2021transbts}, TransBTSv2~\cite{li2023transbtsv2}, UNETR~\cite{hatamizadeh2022unetr}, SwinUNETR~\cite{hatamizadeh2022swin}, and 3DUXNET~\cite{chen2023duxnet}. Diffusion models include DiffBTS~\cite{nie2025diffbts}, Diff-UNet~\cite{xing2023diffusion}, and FCFDiff-Net~\cite{wu2024fcfdiff}. Hybrid architectures include Gate-UNet~\cite{schwehr2024brain}, Optimized U-Net~\cite{futrega2022optimized}, CKD-TransBTS~\cite{lin2023ckd}, MedNeXt~\cite{maani2024advanced}, and OMT-nnU-Net~\cite{zhu2025omt}.

{\scriptsize
\setlength{\tabcolsep}{2.8pt}
\renewcommand{\arraystretch}{1.1}
\begin{longtable}{@{} c c l c c c c c c c c @{}}
\caption{Performance Comparison on BraTS Datasets. HD95 denotes the 95\% Hausdorff Distance.}
\label{tab:brats_all}\\
\toprule
 &  &  & \multicolumn{4}{c}{\textbf{Dice Coefficient}} & \multicolumn{4}{c}{\textbf{HD95 (mm)}} \\
\cmidrule(lr){4-7} \cmidrule(lr){8-11}
\textbf{Dataset} & \textbf{Year} & \textbf{Method} & \textbf{Avg} & \textbf{WT} & \textbf{TC} & \textbf{ET} & \textbf{Avg} & \textbf{WT} & \textbf{TC} & \textbf{ET} \\
\midrule
\endfirsthead

\multicolumn{11}{c}{Table~\ref{tab:brats_all} (continued)}\\
\toprule
 &  &  & \multicolumn{4}{c}{\textbf{Dice Coefficient}} & \multicolumn{4}{c}{\textbf{HD95 (mm)}} \\
\cmidrule(lr){4-7} \cmidrule(lr){8-11}
\textbf{Dataset} & \textbf{Year} & \textbf{Method} & \textbf{Avg} & \textbf{WT} & \textbf{TC} & \textbf{ET} & \textbf{Avg} & \textbf{WT} & \textbf{TC} & \textbf{ET} \\
\midrule
\endhead

\midrule
\multicolumn{11}{r}{\textit{Continued on next page}} \\
\endfoot

\bottomrule
\multicolumn{11}{@{}p{\dimexpr\linewidth-2\tabcolsep}@{}}{%
  \scriptsize
  \textbf{Note:} Some methods show significantly lower Hausdorff Distance metrics. This may stem from differences in calculation methods. Two implementations exist. The first is contour point-based, extracting only boundary points. The second is voxel-based, considering all foreground voxels. Our method adopts the BraTS official voxel-based standard~\cite{bakas2018identifying,karimi2020reducing}.
}\\
\endlastfoot

2020 & 2020 & nnU-Net~\cite{isensee2018nnu} & 0.854 & 0.890 & 0.851 & 0.820 & 14.55 & 8.50 & 17.34 & 17.80 \\
2020 & 2021 & TransBTS~\cite{wang2021transbts} & 0.829 & 0.911 & 0.836 & 0.740 & 3.25 & 3.36 & 2.99 & 3.40 \\
2020 & 2021 & TransUNet~\cite{chen2021transunet} & 0.825 & 0.892 & 0.825 & 0.758 & 3.22 & 3.15 & 2.89 & 3.62 \\
2020 & 2021 & UNETR~\cite{hatamizadeh2022unetr} & 0.816 & 0.902 & 0.813 & 0.732 & 4.90 & 4.31 & 5.74 & 4.64 \\
2020 & 2022 & SwinUNETR~\cite{hatamizadeh2022swin} & 0.831 & 0.917 & 0.826 & 0.749 & 3.89 & 2.86 & 4.31 & 4.50 \\
2020 & 2022 & SwinBTS~\cite{jiang2022swinbts} & 0.823 & 0.891 & 0.804 & 0.774 & 17.06 & 8.56 & 15.78 & 26.84 \\
2020 & 2022 & CKD-TransBTS~\cite{lin2023ckd} & 0.836 & 0.898 & 0.841 & 0.770 & 2.96 & 2.42 & 3.45 & 3.02 \\
2020 & 2022 & FDiff-Fusion~\cite{ding2024denoising} & 0.842 & 0.905 & 0.844 & 0.776 & 2.74 & 2.21 & 3.31 & 2.71 \\
2020 & 2023 & Gate-UNet~\cite{schwehr2024brain} & 0.850 & 0.914 & 0.845 & 0.791 & 3.78 & 2.68 & 4.29 & 4.38 \\
2020 & 2023 & Diff-UNet~\cite{xing2023diffusion} & 0.857 & 0.920 & 0.851 & 0.799 & 2.87 & 1.79 & 3.41 & 3.42 \\
2020 & 2024 & Opt-cNet~\cite{futrega2022optimized} & 0.860 & 0.922 & 0.856 & 0.802 & 2.47 & 1.55 & 2.70 & 3.16 \\
2020 & 2025 & DiffBTS~\cite{nie2025diffbts} & 0.864 & \textbf{0.922} & 0.867 & 0.804 & 2.47 & 1.63 & 2.49 & 3.28 \\
2020 & 2025 & FCFDiff-Net~\cite{wu2024fcfdiff} & 0.854 & 0.916 & 0.860 & 0.786 & 2.36 & 1.92 & 2.57 & 2.58 \\
2020 & 2025 & Ours (Full) & \textbf{0.867} & 0.887 & \textbf{0.876} & \textbf{0.836} & 6.08 & 9.50 & 4.75 & 3.98 \\
2020 & 2025 & Ours (Baseline) & 0.809 & 0.889 & 0.810 & 0.728 & 12.15 & 10.75 & 12.43 & 13.28 \\
\midrule

2021 & 2021 & Ext nnU-Net~\cite{futrega2022optimized} & 0.884 & 0.928 & 0.878 & 0.845 & 10.61 & 3.47 & 7.62 & 20.73 \\
2021 & 2021 & TransBTS~\cite{wang2021transbts} & 0.866 & 0.920 & 0.882 & 0.795 & 8.72 & 4.98 & 4.86 & 16.32 \\
2021 & 2021 & TransUNet~\cite{chen2021transunet} & 0.871 & 0.919 & 0.877 & 0.818 & 8.86 & 6.16 & 7.34 & 13.09 \\
2021 & 2021 & UNETR~\cite{hatamizadeh2022unetr} & 0.854 & 0.890 & 0.847 & 0.825 & 11.14 & 14.42 & 10.22 & 8.79 \\
2021 & 2022 & SwinUNETR~\cite{hatamizadeh2022swin} & 0.890 & 0.926 & 0.885 & 0.858 & 5.21 & 5.83 & 3.77 & 6.02 \\
2021 & 2022 & SwinBTS~\cite{jiang2022swinbts} & 0.866 & 0.918 & 0.848 & 0.832 & 11.40 & 3.65 & 14.51 & 16.03 \\
2021 & 2023 & CKD-TransBTS~\cite{lin2023ckd} & 0.884 & 0.923 & 0.881 & 0.848 & 3.93 & 4.23 & 4.39 & 3.16 \\
2021 & 2024 & Opt-Unet~\cite{futrega2022optimized} & 0.894 & 0.928 & 0.903 & 0.852 & 1.71 & 1.50 & 1.68 & 1.96 \\
2021 & 2024 & Diff-Unet~\cite{xing2023diffusion} & 0.890 & 0.925 & 0.894 & 0.852 & 1.96 & 1.70 & 2.07 & 2.12 \\
2021 & 2024 & Gate-Unet~\cite{schwehr2024brain} & 0.877 & 0.907 & 0.876 & 0.848 & 2.94 & 2.92 & 2.76 & 3.16 \\
2021 & 2025 & FCFDiff-Net~\cite{wu2024fcfdiff} & 0.899 & 0.926 & 0.903 & \textbf{0.869} & 1.86 & 2.16 & 1.83 & 1.58 \\
2021 & 2025 & DiffBTS~\cite{nie2025diffbts} & 0.900 & 0.930 & 0.907 & 0.863 & 1.93 & 2.04 & 1.67 & 2.08 \\
2021 & 2025 & Ours (Full) & \textbf{0.901} & \textbf{0.932} & \textbf{0.912} & 0.861 & 6.81 & 4.19 & 6.79 & 9.44 \\
2021 & 2025 & Ours (Baseline) & 0.848 & 0.860 & 0.872 & 0.811 & 14.07 & 13.49 & 12.40 & 16.33 \\
\midrule

2023 & 2023 & Faking It~\cite{bakas2023brats} & 0.876 & 0.910 & 0.867 & 0.851 & 14.43 & 11.11 & 14.47 & 17.70 \\
2023 & 2023 & MedNeXt+Seg~\cite{maani2024advanced} & 0.871 & 0.906 & 0.863 & 0.843 & 14.06 & 11.70 & 13.10 & 17.37 \\
2023 & 2025 & OMT-nnU-Net~\cite{zhu2025omt} & 0.880 & 0.908 & 0.876 & 0.857 & 12.58 & 11.27 & 12.83 & 13.65 \\
2023 & 2025 & 3D U-Net~\cite{cicek20163d} & 0.787 & 0.855 & 0.805 & 0.702 & 24.63 & 12.50 & 12.90 & 48.50 \\
2023 & 2025 & V-Net~\cite{milletari2016vnet} & 0.767 & 0.859 & 0.778 & 0.645 & 25.43 & 19.00 & 11.50 & 45.80 \\
2023 & 2025 & SegTransVAE~\cite{pham2022segtransvae} & 0.779 & 0.868 & 0.792 & 0.678 & 24.13 & 18.20 & 11.20 & 43.00 \\
2023 & 2025 & SegResNet~\cite{myronenko20183d} & 0.795 & 0.875 & 0.809 & 0.700 & 20.67 & 11.80 & 10.00 & 38.20 \\
2023 & 2025 & Attn U-Net~\cite{oktay2018attention} & 0.812 & 0.889 & 0.823 & 0.725 & 17.73 & 9.80 & 8.70 & 32.70 \\
2023 & 2025 & UNETR~\cite{hatamizadeh2022unetr} & 0.829 & 0.895 & 0.830 & 0.743 & 15.83 & 8.90 & 8.20 & 30.40 \\
2023 & 2025 & SwinUNETR~\cite{hatamizadeh2022swin} & 0.837 & 0.905 & 0.840 & 0.765 & 13.73 & 7.20 & 6.80 & 27.20 \\
2023 & 2025 & 3DUXNET~\cite{chen2023duxnet} & 0.848 & 0.910 & 0.850 & 0.783 & 11.63 & 5.80 & 5.60 & 23.50 \\
2023 & 2025 & TransBTS~\cite{wang2021transbts} & 0.848 & 0.913 & 0.835 & 0.795 & 9.90 & 4.80 & 9.00 & 15.90 \\
2023 & 2025 & TransBTSv2~\cite{li2023transbtsv2} & 0.859 & \textbf{0.918} & 0.860 & 0.800 & 8.00 & 4.00 & 5.00 & 15.00 \\
2023 & 2025 & Ours (Full) & \textbf{0.891} & 0.917 & \textbf{0.899} & \textbf{0.858} & 12.17 & 8.98 & 11.21 & 16.33 \\
2023 & 2025 & Ours (Baseline) & 0.830 & 0.854 & 0.842 & 0.794 & 18.55 & 15.29 & 17.44 & 22.93 \\

\end{longtable}
}

Table~\ref{tab:brats_all} shows that the proposed method achieves average Dice coefficients of 0.8665, 0.9014, and 0.8912 on the three datasets. Compared to CNN methods, the method surpasses nnU-Net by approximately 1.5 percentage points on BraTS 2020. The key advantage is that the cross-modal alignment mechanism establishes long-range dependency modeling through Transformers. Convolution struggles with global spatial relationships. Compared to Transformer methods, the proposed method shows 3 to 8 percentage point improvements. The core advantage lies in systematic integration across three tiers beyond pure visual processing. Compared to diffusion models, better TC and ET performance reflects semantic guidance's advantage over random denoising. TC Dice reaches 0.8758 vs 0.8668, and ET Dice reaches 0.8364 vs 0.8041 on BraTS 2020. Medical concepts transform into spatial attention for targeted enhancement.

The method's core contribution concentrates in the clinically critical ET and TC regions. ET is a small-volume, blurred-boundary target with varied morphology. GPT-4V-generated clinical descriptions combined with CLIP's discriminative representations benefit ET. They enable understanding of concepts like ring enhancement. They transform these concepts into concrete spatial localization. TC requires simultaneous discrimination of necrosis, cystic change, and enhancement. Multi-teacher distillation integrates $T_{\text{MRI}}$, $T_{\text{CT}}$, $T_{\text{US}}$, and $T_{\text{GPT-4V}}$ knowledge. This provides complementary expertise. The WT region shows relatively stable performance. Whole tumor boundaries are clear and large in volume. Most methods achieve good results for WT.

\subsection{Ablation Studies}

\subsubsection{Main Component Ablation}

Table~\ref{tab:ablation_main} reports systematic ablation experiments. We sequentially remove pixel-level multimodal generation, feature-level multi-teacher distillation, and semantic-level CLIP-GPT guidance. This quantifies each component's contribution.

{\scriptsize
\setlength{\tabcolsep}{2.8pt}
\renewcommand{\arraystretch}{1.1}
\begin{longtable}{@{} c c l c c c c c c c c @{}}
\caption{Ablation Study on Main Components}
\label{tab:ablation_main}\\
\toprule
 &  &  & \multicolumn{4}{c}{\textbf{Dice Coefficient}} & \multicolumn{4}{c}{\textbf{HD95 (mm)}} \\
\cmidrule(lr){4-7} \cmidrule(lr){8-11}
\textbf{Dataset} & \textbf{Year} & \textbf{Configuration} & \textbf{Avg} & \textbf{WT} & \textbf{TC} & \textbf{ET} & \textbf{Avg} & \textbf{WT} & \textbf{TC} & \textbf{ET} \\
\midrule
\endfirsthead

\multicolumn{11}{c}{Table~\ref{tab:ablation_main} (continued)}\\
\toprule
 &  &  & \multicolumn{4}{c}{\textbf{Dice Coefficient}} & \multicolumn{4}{c}{\textbf{HD95 (mm)}} \\
\cmidrule(lr){4-7} \cmidrule(lr){8-11}
\textbf{Dataset} & \textbf{Year} & \textbf{Configuration} & \textbf{Avg} & \textbf{WT} & \textbf{TC} & \textbf{ET} & \textbf{Avg} & \textbf{WT} & \textbf{TC} & \textbf{ET} \\
\midrule
\endhead

\midrule
\multicolumn{11}{r}{\textit{Continued on next page}} \\
\endfoot

\bottomrule
\endlastfoot

2020 & 2025 & Full Method & \textbf{0.867} & \textbf{0.887} & \textbf{0.876} & \textbf{0.836} & \textbf{6.08} & 9.50 & \textbf{4.75} & \textbf{3.98} \\
2020 & 2025 & w/o Semantic & 0.845 & 0.889 & 0.844 & 0.802 & 8.68 & \textbf{9.21} & 8.85 & 7.97 \\
2020 & 2025 & w/o Synth CT & 0.848 & 0.863 & 0.864 & 0.817 & 8.30 & 12.45 & 6.05 & 6.39 \\
2020 & 2025 & w/o Sim US & 0.853 & 0.870 & 0.865 & 0.824 & 7.80 & 12.89 & 5.79 & 4.71 \\
2020 & 2025 & w/o Distill & 0.845 & 0.872 & 0.853 & 0.808 & 8.94 & 11.07 & 7.45 & 8.29 \\
2020 & 2025 & Baseline & 0.809 & 0.889 & 0.810 & 0.728 & 12.15 & 10.75 & 12.43 & 13.28 \\
\midrule

2021 & 2025 & Full Method & \textbf{0.901} & \textbf{0.932} & \textbf{0.912} & \textbf{0.861} & \textbf{6.81} & \textbf{4.19} & \textbf{6.79} & \textbf{9.44} \\
2021 & 2025 & w/o Semantic & 0.877 & 0.926 & 0.881 & 0.824 & 9.91 & 5.13 & 9.24 & 15.37 \\
2021 & 2025 & w/o Synth CT & 0.871 & 0.887 & 0.889 & 0.837 & 11.99 & 11.35 & 11.24 & 13.38 \\
2021 & 2025 & w/o Sim US & 0.880 & 0.897 & 0.897 & 0.846 & 9.89 & 8.97 & 9.02 & 11.66 \\
2021 & 2025 & w/o Distill & 0.877 & 0.920 & 0.884 & 0.827 & 10.25 & 6.13 & 10.13 & 14.50 \\
2021 & 2025 & Baseline & 0.848 & 0.860 & 0.872 & 0.811 & 14.07 & 13.49 & 12.40 & 16.33 \\
\midrule

2023 & 2025 & Full Method & \textbf{0.891} & \textbf{0.917} & \textbf{0.899} & \textbf{0.858} & \textbf{12.17} & \textbf{8.98} & \textbf{11.21} & \textbf{16.33} \\
2023 & 2025 & w/o Semantic & 0.871 & 0.916 & 0.868 & 0.830 & 14.44 & 9.29 & 14.78 & 19.24 \\
2023 & 2025 & w/o Synth CT & 0.871 & 0.888 & 0.882 & 0.841 & 13.86 & 11.33 & 12.89 & 17.35 \\
2023 & 2025 & w/o Sim US & 0.877 & 0.894 & 0.890 & 0.848 & 13.12 & 10.24 & 12.13 & 16.98 \\
2023 & 2025 & w/o Distill & 0.865 & 0.893 & 0.867 & 0.835 & 14.57 & 10.30 & 14.98 & 18.44 \\
2023 & 2025 & Baseline & 0.830 & 0.854 & 0.842 & 0.794 & 18.55 & 15.29 & 17.44 & 22.93 \\

\end{longtable}
}

The complete method improves average Dice by 5.37 to 6.12 percentage points over baseline. It improves Hausdorff Distance by 34.4\% to 51.6\%. The three tiers show differentiated functional division. Removing semantic guidance causes 1.99 to 2.44 percentage point drops. ET is most severely affected with 2.81 to 3.75 point drops. This reveals semantic guidance's core value for high-complexity tasks. Abstract concepts provide prior knowledge unavailable to pure visual features. Synthetic CT removal causes 1.85 to 3.04 point drops. ET is most affected with 1.66 to 2.44 point drops. Density information identifies heterogeneous structures like calcification and hemorrhage. Simulated ultrasound contributes 1.34 to 2.16 point improvements. Hausdorff Distance deterioration (3.98 mm vs 4.71 mm on BraTS 2020) proves physics-driven ultrasound's value. Strong echo features aid boundary refinement. Feature-level distillation causes 2.20 to 2.60 point drops. TC is most damaged with 2.25 to 3.12 point drops. Tumor core discrimination requires integrating complementary knowledge from multiple imaging mechanisms.

\subsubsection{Semantic-Level Internal Mechanism Ablation}

Table~\ref{tab:ablation_semantic} focuses on semantic-level internal mechanisms. We separately remove multi-view guidance, cross-modal fusion, and semantic attention generation.

{\scriptsize
\setlength{\tabcolsep}{2.8pt}
\renewcommand{\arraystretch}{1.1}
\begin{longtable}{@{} c c l c c c c c c c c @{}}
\caption{Semantic-Level Mechanism Ablation}
\label{tab:ablation_semantic}\\
\toprule
 &  &  & \multicolumn{4}{c}{\textbf{Dice Coefficient}} & \multicolumn{4}{c}{\textbf{HD95 (mm)}} \\
\cmidrule(lr){4-7} \cmidrule(lr){8-11}
\textbf{Dataset} & \textbf{Year} & \textbf{Method} & \textbf{Avg} & \textbf{WT} & \textbf{TC} & \textbf{ET} & \textbf{Avg} & \textbf{WT} & \textbf{TC} & \textbf{ET} \\
\midrule
\endfirsthead
\multicolumn{11}{c}{Table~\ref{tab:ablation_semantic} (continued)}\\
\toprule
 &  &  & \multicolumn{4}{c}{\textbf{Dice Coefficient}} & \multicolumn{4}{c}{\textbf{HD95 (mm)}} \\
\cmidrule(lr){4-7} \cmidrule(lr){8-11}
\textbf{Dataset} & \textbf{Year} & \textbf{Method} & \textbf{Avg} & \textbf{WT} & \textbf{TC} & \textbf{ET} & \textbf{Avg} & \textbf{WT} & \textbf{TC} & \textbf{ET} \\
\midrule
\endhead
\midrule \multicolumn{11}{r}{\textit{Continued on next page}}
\endfoot
\bottomrule
\endlastfoot

2020 & 2025 & Full Method & \textbf{0.867} & \textbf{0.887} & \textbf{0.876} & \textbf{0.836} & \textbf{6.08} & 9.50 & \textbf{4.75} & \textbf{3.98} \\
2020 & 2025 & w/o Multi-view & 0.857 & 0.890 & 0.864 & 0.818 & 7.27 & 9.32 & 6.04 & 6.45 \\
2020 & 2025 & w/o Sem Fusion & 0.860 & 0.890 & 0.868 & 0.821 & 6.98 & \textbf{9.27} & 6.00 & 5.67 \\
2020 & 2025 & w/o Sem Attn & 0.855 & 0.889 & 0.857 & 0.820 & 7.42 & 9.55 & 7.03 & 5.67 \\
\midrule

2021 & 2025 & Full Method & \textbf{0.901} & \textbf{0.932} & \textbf{0.912} & \textbf{0.861} & \textbf{6.81} & \textbf{4.19} & \textbf{6.79} & \textbf{9.44} \\
2021 & 2025 & w/o Multi-view & 0.887 & 0.926 & 0.894 & 0.842 & 8.89 & 5.14 & 9.33 & 12.20 \\
2021 & 2025 & w/o Sem Fusion & 0.889 & 0.930 & 0.891 & 0.845 & 8.49 & 4.24 & 9.35 & 11.87 \\
2021 & 2025 & w/o Sem Attn & 0.885 & 0.925 & 0.888 & 0.841 & 9.36 & 5.33 & 9.92 & 12.83 \\
\midrule

2023 & 2025 & Full Method & \textbf{0.891} & \textbf{0.917} & \textbf{0.899} & \textbf{0.858} & \textbf{12.17} & \textbf{8.98} & \textbf{11.21} & \textbf{16.33} \\
2023 & 2025 & w/o Multi-view & 0.878 & 0.916 & 0.880 & 0.840 & 13.53 & 9.12 & 13.33 & 18.13 \\
2023 & 2025 & w/o Sem Fusion & 0.881 & 0.920 & 0.881 & 0.840 & 13.14 & 8.73 & 12.98 & 17.71 \\
2023 & 2025 & w/o Sem Attn & 0.873 & 0.914 & 0.872 & 0.832 & 13.95 & 9.39 & 13.72 & 18.75 \\

\end{longtable}
}

The three sub-modules form a layer-by-layer progressive mapping chain. Multi-view guidance resolves the 2D-3D dimensional gap with 0.93 to 1.40 point drops when removed. Single-perspective slices cannot represent 3D tumor structures. Complementary information from orthogonal planes forms global understanding. This is crucial for identifying infiltrative growth and spatial heterogeneity. Cross-modal fusion enables concept-feature alignment with 0.67 to 1.27 point drops when removed. It maps radiological descriptions to the same semantic space as visual features. Gating dynamically determines reliance on visual versus textual information. Semantic attention generation shows most pronounced drops when removed. These include 1.11 to 1.86 points overall and 1.60 to 2.56 points for ET. Hausdorff Distance increases from 9.44 mm to 12.83 mm on BraTS 2021. This reveals its critical role in transforming abstract concepts into explicit spatial weight distributions. It achieves the conversion from understanding what the tumor is to knowing where the tumor is.

\subsection{Visualization Analysis}

\begin{figure}[htbp]
  \centering
  \includegraphics[width=0.8\textwidth,keepaspectratio]{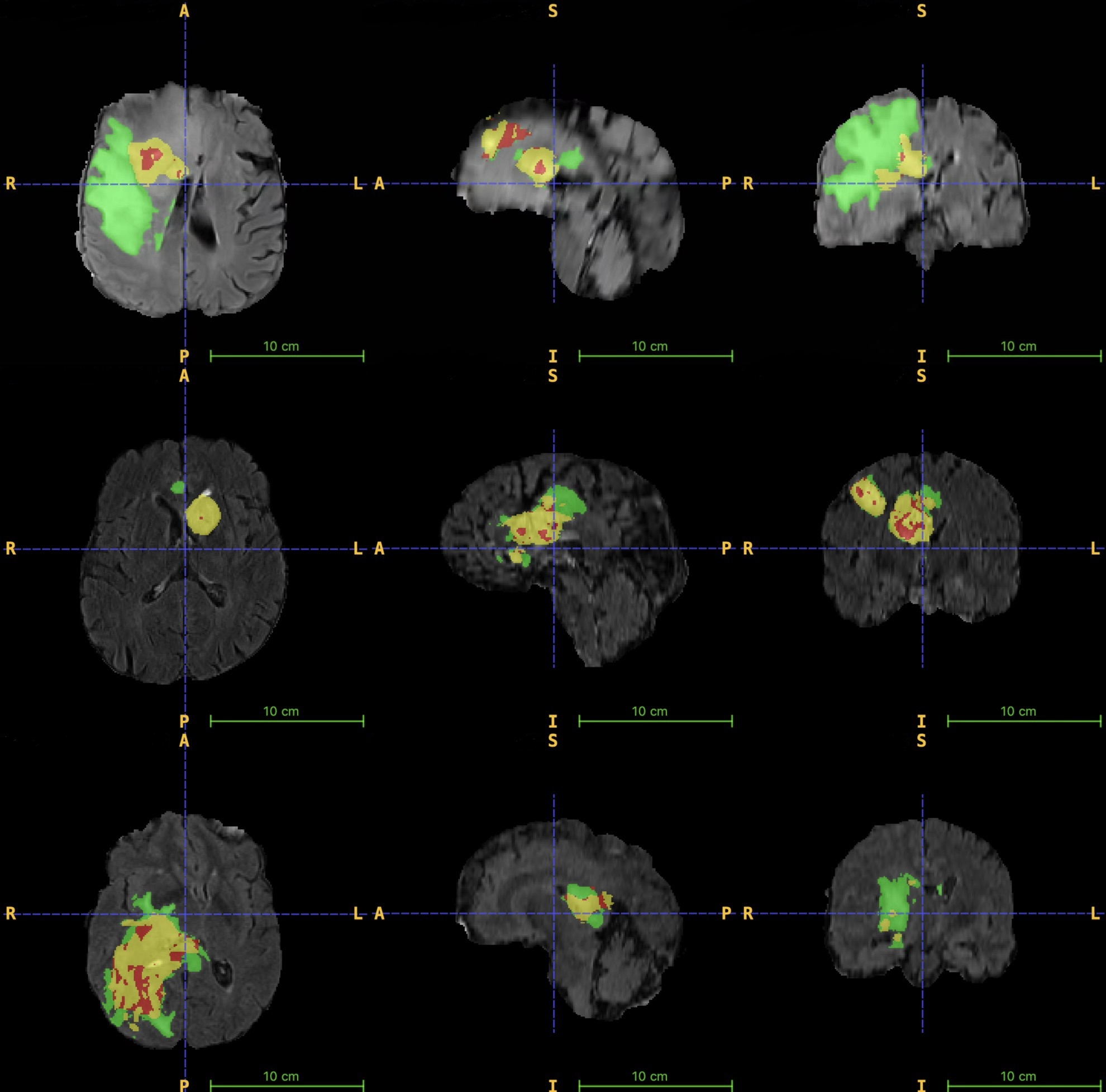}
  \caption{Three-dimensional visualization based on ITK-SNAP: grayscale MRI with whole tumor (green), tumor core (yellow), enhancing tumor (red) masks displayed in axial, sagittal, and coronal views.}
  \label{fig:itksnap-3d}
\end{figure}

Figure~\ref{fig:itksnap-3d} shows ITK-SNAP~\cite{yushkevich2006user} 3D visualization. WT (green), TC (yellow), and ET (red) overlay on grayscale MRI across three orthogonal views. The red ET exhibits ring enhancement with irregular boundaries. Yellow TC surrounds it. Green WT extends finger-like along white matter tracts. This suggests infiltrative edema spreading along sulci and periventricular directions. The three-dimensional coherence and boundary consistency align with quantitative trends. ET and TC Dice improve while Hausdorff Distance reduces. This demonstrates that pixel-level multimodal generation provides boundary and density priors for sharper contours. Feature-level alignment reduces mis-segmentation at tissue junctions. Semantic-level CLIP guidance transforms concepts into spatial attention. This is effective for small-volume ET regions.

\subsection{Clinical Interpretability: Semantic Attention Case Study}

\begin{figure}[htbp]
  \centering
  \includegraphics[width=\textwidth,keepaspectratio]{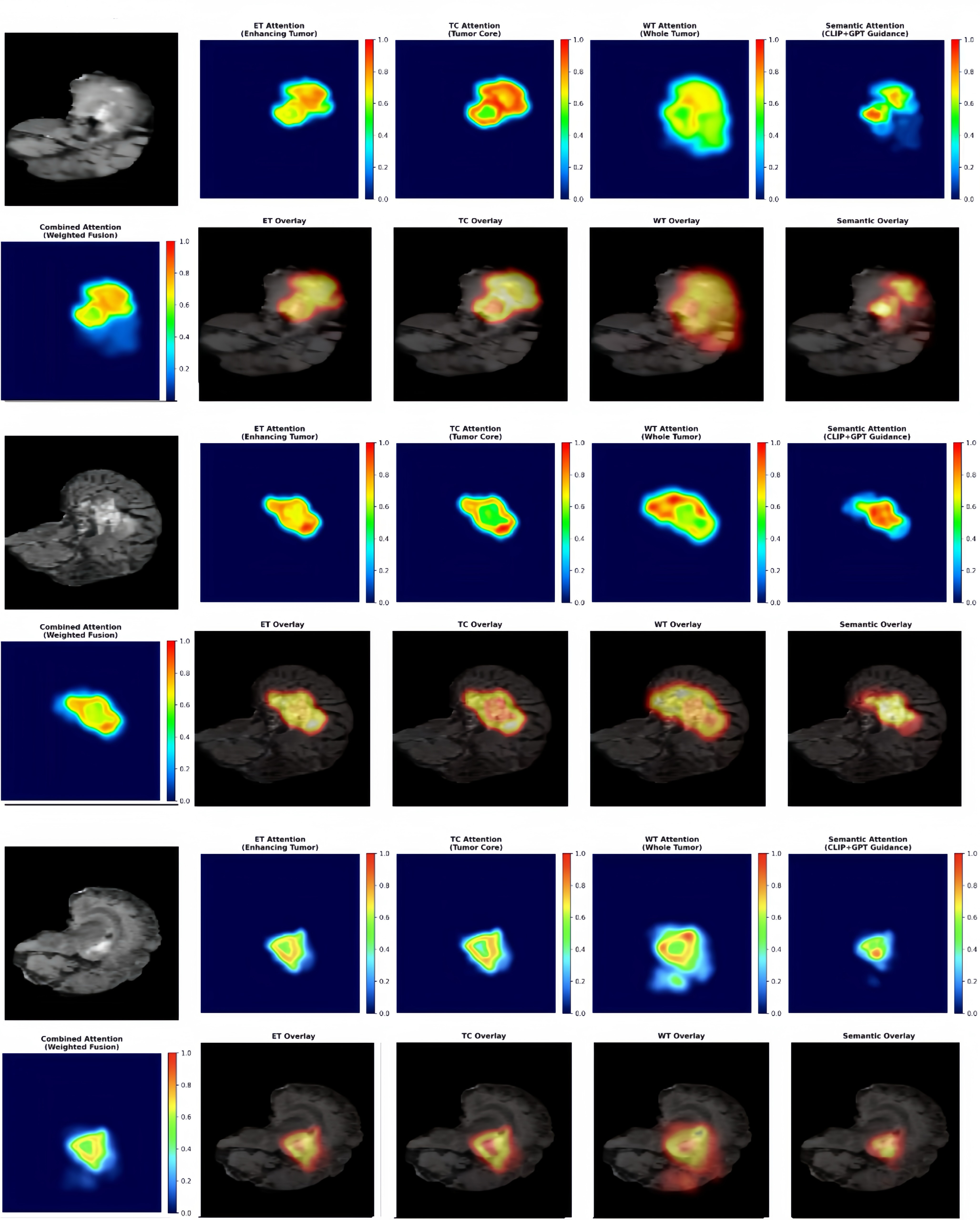}
  \caption{Semantic attention visualization: left column shows grayscale MRI; next three columns are enhancing tumor, tumor core, whole tumor dedicated attention heatmaps; rightmost column shows semantic attention (CLIP and GPT guided); second row gives thermal overlay with base image, with leftmost being weighted fusion attention.}
  \label{fig:semantic-attn}
\end{figure}

Figure~\ref{fig:semantic-attn} visualizes semantic attention distribution. ET attention forms closed high-value bands at enhancement ring edges with suppressed centers. This is consistent with ring enhancement and necrotic core radiological patterns. TC attention extends inward and outward. It covers necrotic and non-enhancing parenchyma with smoother boundaries. WT attention extends along FLAIR high signals. It shows finger-like spread along white matter tracts. This reflects infiltrative edema spatial trajectories. CLIP and GPT-guided semantic attention increases weights at blurred boundaries or irregular morphologies. It maintains low responses to non-tumor structures like ventricles and vessels. This suppresses false detections. The weighted fused heatmap gives peak values in multi-head consensus regions. It shows continuous boundaries and reduced speckle artifacts. This improves focus and recall of small-volume ET. This visualization provides surgical and radiotherapy reference. The ET head suggests active tumor margins. The TC head reveals core extent. The WT head reflects edema expansion. Semantic attention provides evidence for why enhance or why suppress. This helps radiologists quickly locate suspicious regions for manual corrections.

\subsection{Computational Efficiency Analysis}

Table~\ref{tab:efficiency} compares the complete method and ablation variants. It examines computational complexity, inference speed, and resource consumption. This analyzes the accuracy-cost trade-off.

{\small
\setlength{\tabcolsep}{4pt}
\renewcommand{\arraystretch}{1.1}
\begin{longtable}{@{} l r r r r @{}}
\caption{Computational Efficiency Analysis}
\label{tab:efficiency}\\
\toprule
\textbf{Configuration} & \textbf{Params (M)} & \textbf{FLOPs (G)} & \textbf{Time (ms)} & \textbf{Throughput (s/s)} \\
\midrule
\endfirsthead

\multicolumn{5}{c}{Table~\ref{tab:efficiency} (continued)}\\
\toprule
\textbf{Configuration} & \textbf{Params (M)} & \textbf{FLOPs (G)} & \textbf{Time (ms)} & \textbf{Throughput (s/s)} \\
\midrule
\endhead

\midrule
\multicolumn{5}{r}{\textit{Continued on next page}} \\
\endfoot

\bottomrule
\endlastfoot

Baseline & 29.55 & 575.89 & 40.94 & 24.42 \\
\midrule
w/o Synth CT & 184.48 & 738.88 & 84.09 & 11.89 \\
w/o Sim US & 184.54 & 766.35 & 84.36 & 11.85 \\
w/o Distill & 183.72 & 760.17 & 81.79 & 12.23 \\
w/o Semantic & 31.36 & 747.69 & 65.10 & 15.36 \\
w/o Multi-view & 31.36 & 747.69 & 65.17 & 15.35 \\
w/o Sem Fusion & 183.66 & 661.98 & 73.45 & 13.61 \\
w/o Sem Attn & 184.30 & 774.41 & 85.53 & 11.69 \\
\midrule
Full Method & 184.55 & 774.67 & 88.12 & 11.35 \\

\end{longtable}
}

The complete method shows significant increases in parameters, computation, and inference time compared to baseline. However, this computational overhead yields substantial performance improvements. Average Dice improves by 5.37 to 6.12 percentage points across three datasets. The clinically critical ET region improves by 8.1\% to 14.9\%. Hausdorff Distance improves by 28.8\% to 70.0\%. Component-wise analysis shows reasonable accuracy-cost trade-offs. Pixel-level generation occupies only 4\% to 5\% of inference time. Yet it brings 1.34 to 1.85 point Dice improvements. Multi-source data's contribution far exceeds computational cost. Feature-level distillation contributes 2.20 points in TC with 6.39 ms additional time. This is important for tissue type discrimination. Semantic-level fusion accounts for 82.9\% of parameters. Removing its sub-modules causes 1.60 to 2.56 point ET Dice drops. This proves its irreplaceability in transforming concepts to spatial attention. From a clinical perspective, surgical planning and radiotherapy require much higher segmentation accuracy than speed. The 88.12 ms inference time is completely acceptable. Nearly 15\% accuracy improvement in key regions directly affects treatment decision accuracy.

\section{Discussion}

The proposed method shows more obvious advantages in ET and TC regions compared to WT. This is highly consistent with the design philosophy of the three-tier fusion architecture. WT is a large-volume region with clear boundaries. Most deep learning methods can already achieve good segmentation for WT. This leaves limited room for improvement. In contrast, the ET region has small volume, blurred boundaries, and varied morphology. Traditional pure visual methods struggle to accurately identify ET. The semantic guidance mechanism improves small target recognition. It transforms clinical concepts like ring enhancement and irregular boundaries into spatial attention. This transformation uses FiLM and attention generation. The TC region requires simultaneous discrimination of multiple tissue types. These include necrosis, cystic change, and enhancement. Multi-teacher distillation integrates complementary multimodal knowledge to form synergy. Comparison with diffusion models~\cite{qin2024btsegdiff} shows that the proposed method achieves better performance in small target regions. Semantic attention localization provides this advantage. It reveals that semantic guidance surpasses random denoising processes when handling irregular blurred boundaries.

Despite achieving good performance improvements, the three-tier fusion architecture still has room for improvement. Pixel-level multimodal generation, feature-level multi-teacher distillation, and semantic-level GPT-4V calls increase inference time by approximately 115\%. Future work can optimize this through lightweight networks, knowledge compression, and efficient semantic extraction. Additionally, the proposed method performs consistently across three BraTS datasets. However, all target gliomas as a specific disease. Cross-disease generalization ability remains to be verified. Medical text descriptions at the semantic level need redesign for different diseases. Future work can explore more general medical concept ontologies and cross-disease semantic transfer learning. This would apply the three-tier fusion concept to a broader range of medical imaging tasks.

\section{Conclusion}

This paper proposes a three-tier fusion architecture for brain tumor segmentation. It systematically integrates pixel-level multimodal generation, feature-level cross-modal alignment, and semantic-level concept guidance. Validation on the BraTS 2020, 2021, and 2023 datasets demonstrates substantial improvements in segmentation accuracy. This is particularly true for clinically critical small-volume targets such as enhancing tumors. Ablation studies confirm the synergistic contributions of each architectural component. They also confirm the unique effectiveness of semantic-level CLIP-GPT guidance for complex target localization. The proposed architecture establishes a systematic pathway for integrating clinical knowledge with deep learning-based segmentation. It bridges the gap between data-driven learning and domain expertise in medical imaging. Computational efficiency and cross-disease generalization remain areas for future optimization. However, the method's accuracy gains in key tumor regions provide meaningful clinical value for diagnosis and treatment planning. This demonstrates the potential of multi-tier fusion strategies in advancing medical image analysis.

\section*{Data Availability}

The datasets used in this study are publicly available through the BraTS Challenge:

\begin{itemize}
\item BraTS 2020: \url{https://www.med.upenn.edu/cbica/brats2020/data.html}
\item BraTS 2021: \url{https://www.synapse.org/Synapse:syn25829067}
\item BraTS 2023: \url{https://www.synapse.org/Synapse:syn51156910}
\end{itemize}

All datasets are available for research purposes upon registration and acceptance of the data usage agreement. Code will be available upon publication acceptance.

\section*{CRediT Author Statement}

Mingda Zhang: Conceptualization, Methodology, Software, Validation, Formal analysis, Investigation, Resources, Data Curation, Writing -- Original Draft, Writing -- Review \& Editing, Visualization, Project administration.

\section*{Declaration of Competing Interest}

The author declares no competing interests.

\section*{Acknowledgments}

This research received no specific grant from funding agencies.

\end{document}